# What does it mean to "make sense" of physics?

**Tor Ole B. Odden[1]**

[1] Center for Computing in Science Education, Department of Physics, University of Oslo, Oslo, Norway

## Introduction

What does it mean to "make sense" of physics? It's not a simple question. Most people have an intuitive feeling for when things do (or do not) make sense to them. But, putting this feeling into words—especially actionable words—is another task entirely.

Generally speaking, there are two ways in which we commonly use the term "make sense."[1] The first connotes reasonability: for example, "it makes sense to start saving for college early." Of the two, this is likely the more familiar meaning for the average physics teacher, since evaluating reasonability is a habit that we commonly teach in physics problem-solving. We regularly instruct our students to check if their answers "make sense," and there are numerous dependable strategies for this kind of evaluative sensemaking such as unit checks, limiting-case analysis, order-of-magnitude estimation, and qualitatively examining relationships between variables.[2]

The second way we use "make sense" connotes understanding, intuition, and explainability (or lack thereof): for example "it makes no sense to me how something as heavy as an airplane is able to fly." This meaning is harder to put into action since it tends to be much more intuitive and emotional—what feels sensible to one person may be incoherent to someone else. I call this "intuitive sensemaking," to distinguish it from the evaluative sensemaking described above. It is this meaning I would like to unpack here.

## What does it mean to make intuitive sense of physics?

One of the best descriptions I have ever found of this kind of intuitive sensemaking comes from Seymour Papert, the MIT mathematician, computer scientist, and educator, who describes it as follows in his seminal book *Mindstorms*[3]:

> Everyone knows the unpleasant feeling evoked by running into a counterintuitive phenomenon where we are forced, by observation or by reason, to acknowledge that reality does not fit our expectations. Many people have this feeling when faced with the perpetual motion of a Newtonian particle, with the way a rudder turns a boat, or with the strange behavior of a toy gyroscope. In all these cases intuition seems to betray us. Sometimes there is a simple "fix"; we see that we made a superficial mistake. But the interesting cases are those where the conflict remains obstinately in place however much we ponder the problem. These are the cases where we are tempted to conclude that "intuition cannot be trusted." In these situations we need to improve our intuition, to debug it, but the pressure on us is to abandon intuition and rely on equations instead. Usually when a student in this plight goes to the physics teacher saying, "I think the gyroscope should fall instead of standing upright," the teacher responds by writing an equation to *prove* that the thing stands upright. But that is not what the student needed. He already *knew* that it would stay upright, and this knowledge hurt by conflicting with intuition. By *proving* that it will stand upright the teacher rubs salt in the wound but does nothing to heal it. What the student needs is something quite different: better understanding of himself, not of the gyroscope. He wants to know why his intuition gave him a wrong expectation. He needs to know how to work on his intuitions in order to change them. (p. 144)

There is a lot to unpack in this quote. First, Papert argues, making sense involves stumbling on a phenomenon that conflicts with one's intuition—something that doesn't "fit" with the way we think the world works[4]. This phenomenon, Papert specifies, can either be something physical and real, like a gyroscope, or an abstracted model that students have been asked to accept, like a Newtonian particle. In some cases, the inconsistency is easy to resolve. However, those cases require little or no sensemaking, since they are easily resolved "by inspection." The cases that do require sensemaking are those that remain stubbornly unresolved no matter how much we turn over the perceived inconsistencies within our heads.

Sometimes, this feeling of disconnect can lead to a lot of frustration. In search of a fix, we might even be tempted give up on physical intuition entirely. In a physics context, this option is extra tempting since there is always the possibility of substituting formal, mathematical arguments for physical intuition. However, Papert argues that this is the wrong approach. In the lingo of a computer scientist, our intuition needs debugging; if we give up before the bug has been fixed, we set ourselves up for more trouble further down the line.

So, Papert offers the following advice: what we really need is better understanding of ourselves—that is, an understanding why, exactly, this phenomenon feels so inconsistent with the way we think the world should work. Coming to this kind of understanding may take a great deal of metacognitive reflection, but it is ultimately worthwhile; just as identifying the location and nature of a bug in one's code is often most of the job of fixing it, being able to articulate what it is that doesn't make sense is often half the battle (or more!) of resolving this kind of inconsistency.

In my research, I have seen this struggle firsthand. Students are often vexed—even incensed—when they realize that their intuitions do not fit with their observations.[5] Nailing down what, exactly, it is that bothers them often takes extensive discussion, requiring them to revisit the problem from many different conceptual angles and try on many possible explanations in the process of debugging their intuitions.[6] However, once they have done the intellectual work to diagnose their own thinking, the path to resolution often becomes much more tractable (as discussed in the next section), even if it is not necessarily straightforward.

Here, however, Papert adds a caveat for those of us with the responsibility of teaching physics: do not be tempted to simply reiterate the existence of the counterintuitive phenomenon. In other words, when a student says "it doesn't make sense how a gyroscope can stand up like that!" we, as teachers, should resist the impulse to immediately pull out a mathematical proof or derivation. Not only is this unhelpful—the student already knows *that* something counterintuitive

exists—but it is actually counterproductive since it both exacerbates the student's discomfort and implicitly reiterates the message that a student should "abandon intuition and rely on equations instead." Rather, our job is to work *with* the student—to listen, to explain, and to help the student diagnose their own thinking.

### How do we help students resolve intuitive inconsistencies?

Once a student has articulated the conflict in their intuition, there are numerous strategies that a teacher might use to help them resolve it. For example, one might use an analogy that compares the counterintuitive phenomenon to something more familiar, like the classic analogy between the behavior of an electrical circuit and fluid moving through pipes. One might chain together several analogies into a kind of "conceptual blend", like describing the behavior of an E-M wave by combining aspects of pressure-based sound waves and transverse string waves.[7] One might explain a counterintuitive macroscopic phenomenon by connecting it to more intuitive microscopic phenomena, like explaining how the normal force is caused by the cumulative pushback of many spring-like interatomic bonds.[8] One might help the student build a "causal story" or chain of cause-and-effect, so that they can explain how the phenomenon came about in the first place.[9] Or, one might help the student reframe or recategorize an idea, for example pointing out that even though we talk about heat as if it is a substance it is really more of a flow of energy.[10] In all cases, however, the goal is to help a student connect and situate the counter-intuitive phenomenon within their broader intuitions for how the world works and show that the two are coherent with one another.[11]

One might also help a student look at a problem from several different conceptual angles, in case what they are looking for is simply a different perspective on the phenomenon. For example, consider the following problem:

> *If you take an idealized, charged parallel-plate capacitor, unhook it from any other circuit or electrical components, and physically pull apart the plates, what happens to the voltage difference across the capacitor?*

The answer is that the voltage difference increases. Many students find this answer to be strange, unintuitive, or unsatisfactory and it is not difficult to see why—after all, electric field and electric potential both decrease in magnitude as you move away from a charged source. Yet, here, when you move two charged plates apart, the electric potential difference *increases*. How can one explain this?

There are several possible solutions to this problem:

1. Because the plates are isolated from one another, the charge on each plate is trapped. So, it follows that the net charge on each plate, q, stays constant as you pull them apart. Thus,
$$C = \frac{q}{\Delta V} = \varepsilon_0 \varepsilon_r \frac{A}{d} \tag{1}$$
(where C is the capacitance, A is the area of the plates, and d is the separation) and the voltage difference increases in proportion to the distance between the plates.

2. Because we are dealing with an idealized parallel-plate capacitor, the electric field, E, will be constant between the plates (at least for small displacements). Then, using the

44relationship between electric field and electric potential for a constant E:
$$\Delta V = Ed \tag{2}$$
implying that voltage difference increases with plate separation.

3. From a work/energy perspective, you (the external force) are injecting energy into the system by doing work on the plates. This energy must go somewhere—in this case, it goes into increasing the electric potential energy between them, which increases the voltage difference across the capacitor.

4. You can create an analogy between electric potential and gravitational potential in which electric potential acts as "height" for charges. Thus, positive charges make a kind of "hill" around themselves and negative charges create a corresponding "pit." When the plates are close together, the pit around it subsumes a bit of the hill (reducing its overall height) and the hill fills in a bit of the pit (reducing its depth). When you draw the two plates apart, this overlap is reduced, and the difference in "height" between the top of the hill and the bottom of the pit increases. Therefore, the potential difference (height difference) increases.

By examining (and discussing) a problem from these different angles, we increase the likelihood that we will be able to understand what might be bothering a student. And, we increase the range of possible responses and explanations we can deploy, beyond just a single mathematical derivation or proof.

Here again, however, Papert offers an implicit warning—despite our best efforts, there is no guarantee that even the best constructed explanation will necessarily resolve a student's difficulties. This is because the connections needed to resolve an inconsistency are as individual as the students themselves. In other words, although a teacher can help a student in the process—providing guidance, ideas, examples, and resources—it's up to the student to do the hard work of debugging their own knowledge. No one else can do it for them. In the words of Eleanor Duckworth[12], "Thoughts are our way of connecting things up for ourselves. If others tell us about the connections they have made, we can only understand them to the extent that we do the work of making these connections ourselves... there is no way to guarantee that the same words will cue in the same way for every child." (p. 26-27). However, with that said, even if a student does not manage to resolve their difficulty or arrive at the answer given in their textbook, the process of sensemaking itself is still valuable—although they may be frustrated, such students will emerge with a greater understanding of themselves and their own intuitions, providing a solid foundation for future learning.

Thus, as teachers we can help students make intuitive sense of physics by listening before we speak; by trying to understand, as best we can, the students' perspectives and ideas; and by helping students to debug, heal, and strengthen their intuitions and understandings of the world.

## Acknowledgments

I would like to thank Rosemary Russ, Elizabeth Gire, Eric Kuo, and John Burk for many thoughtful conversations that led to the ideas in this paper.